\documentclass[10pt,twocolumn,twoside]{IEEEtran}
\usepackage{graphicx}
\usepackage{amsmath}
\usepackage{epsfig}
\usepackage{tikz}
\usepackage{pgfplots}
\usepackage{psfrag}
\usepackage{amssymb}
\usepackage{bm}
\usepackage{amsthm}
\usepackage{color}
\usepackage{citesort}
\usepackage{algorithm}
\usepackage{algpseudocode}
\usepackage{pifont}

\title{Securing Relay Networks With Artificial Noise: An Error Performance Based Approach}

\author{
        \IEEEauthorblockN{Ying Liu\IEEEauthorrefmark{2}$^{,\, *}$, Liang Li\IEEEauthorrefmark{2}, George C. Alexandropoulos\IEEEauthorrefmark{3}, and Marius Pesavento\IEEEauthorrefmark{2}}

    \IEEEauthorblockA{\IEEEauthorrefmark{2}Communication Systems Group, Technische Universit\"at Darmstadt, Darmstadt, Germany
    \\\{yliu, lli, pesavento\}@nt.tu-darmstadt.de}

    \IEEEauthorblockA{\IEEEauthorrefmark{3}Mathematical and Algorithmic Sciences Lab,
France Research Center, Huawei Technologies Co. Ltd.
    \\george.alexandropoulos@huawei.com}

\thanks{Ying Liu$^{*}$ (ying.liu@nt.tu-darmstadt.de) is the Corresponding Author.
This work was presented in part in IEEE 8th Sensor Array and Multichannel Signal Processing Workshop (SAM), 22-25 Jun. 2014, A Coru\~{n}a \cite{LiuSAM14}.}
\thanks{Y. Liu, L. Li, and M. Pesavento acknowledge the financial support of the Seventh Framework Programme for Research of the European Commission under grant number: ADEL-619647. This work was partially performed within the LOEWE Priority Program Cocoon (http://www.cocoon.tu-darmstadt.de) supported by the LOEWE research initiative of the state of Hesse/Germany. }
}

\begin{document}
\maketitle

\begin{abstract}

We apply the concept of artificial and controlled interference in a triangular relay network with an untrusted relay aiming at enhancing the wireless communication secrecy between the source and the destination node. In order to shield the square quadrature amplitude modulated (QAM) signals transmitted from the source node to the relay, the destination node designs and transmits artificial noise (AN) symbols to jam the relay reception. The objective of our considered AN design is to degrade the error probability performance at the untrusted relay, for different types of channel state information (CSI) at the destination. By considering perfect knowledge of the instantaneous CSI of the source-to-relay and relay-to-destination links, we first present an analytical expression for the symbol error rate (SER) performance at the relay. Based on the assumption of an average power constraint at the destination node, we then derive the optimal phase and power distribution of the AN that maximizes the SER at the relay. Furthermore, we obtain the optimal AN design for the case where only statistical CSI is available at the destination node. For both cases, our study reveals that the Gaussian distribution is generally not optimal to generate AN symbols. In particular, the conducted numerical results show that the Gaussian distributed AN is far from optimal. The presented AN design takes into account practical parameters for the communication links, such as QAM signaling and maximum likelihood decoding.

\end{abstract}

\section{Introduction}

\IEEEPARstart{T}{o} date, secure data communications heavily rely on modern cryptography.
{Since Diffie and Hellman \cite{DiffieIT76} first proposed a key exchange protocol based on computational intractability, computational cryptography approaches have been extensively studied, such as the popular one based on the integer factorization problem \cite{ShorSIAM97}.}
However, the emergence of high-performance computers may challenge the existing cryptographic algorithms relying on computational hardness.
As a complementary strategy to provide secure data communications, physical layer secrecy has recently drawn considerable attention.
{In particular, physical layer secrecy is viewed as a promising solution to provide wireless secrecy in 5G, since it does not depend on computational complexity, and has a high scalability to allow the coexistence of communication terminals with different levels of hierarchical architectures \cite{YangTranMag15}.
What's more, physical l ayer secrecy can either provide direct secure data communication or assist the distribution of cryptographic keys, which makes it particularly favorable in 5G networks \cite{YangTranMag15}.
Since preliminary works \cite{Wyner75,CsiszarIT78} characterized the secrecy capacity for wiretap channels, secrecy communication has been extensively studied for various channel models and network setups, such as single-hop wiretap channels \cite{Liang08,ZhouTVT10,KhistiIT10,KwanNgTWC11,OggierIT11,LiuIT13}, multi-user networks \cite{MukherjeeAllerton09,GeraciTcom11}, and relay networks  \cite{TekinIT08,DongTSP10,MukherjeeSPAWC10,FakoorianTSP11,MukherjeeTSP13}.
{In relay networks, secrecy is an important issue even without the presence of external eavesdroppers, since it is often desirable for the source to keep its messages confidential from the relay, despite the fact that the relay accedes to the request from the source to help forward its messages to the destination \cite{OohamaITW01,OohamaISIT07,HeIT10,HuangTSP13,WangWCL14}.
After first proposed in \cite{OohamaITW01}, this untrusted relay scenario has drawn considerable attention, since it finds diverse and important applications in modern communication systems.
For example, in heterogeneous networks, the relay terminal may have a lower secrecy clearance than the source and destination pair \cite{HeIT10,HuangTSP13}. Therefore, the relay can be partly malicious in the sense that it still functions in compliance with the relaying protocol, whilst it leaks secret information.
Another application is the multiple level access control in wireless sensor networks, where sensors have different authorizations and sensitivities depending on their roles and collected data \cite{WangIEEETUT06,PanjaCC08,LeeINSS10,RohokaleJCyber12}.
In such setups, the relaying node might be only allowed to help forwarding messages from one terminal to another, as not all terminals have direct access to other members in the sensor group.
For instance, in a secure sensor network at an airport, terminals with low secrecy levels can access limited data, whereas a few terminals at a much higher secrecy level are allowed to access all the data \cite{PanjaCC08}.
}


To shield the messages from the untrusted relay and enhance the secrecy of the wireless communication between the source and destination, one popular scheme at the physical layer is to introduce controlled artificial noise (AN) to efficiently jam the signal reception at the relay.
This technique has recently been studied from an information-theoretic perspective, such as secrecy rate and secrecy outage probability, see e.g., \cite{Jeong,SunTVT12,MoWCNC13,HuangTSP13,VishwakarmaICC13,PeiTSP14}.
However, such metrics are in general valid for ideal communication assumptions of continuous input messages and random encoding schemes with asymptotically large block lengths.
In order to take discrete modulation alphabets and finite block lengths into consideration, other secrecy performance metrics have also been proposed, such as bit error rate \cite{KlincTIFS11,MukherjeeCL13}.
The observation that in modern wireless communication systems square quadrature amplitude modulation (QAM) is widely used motivates us to address the problem of how to optimally apply AN to enhance the physical layer secrecy in an untrusted relay network.
The symbol error rate (SER) of the demodulated signal at the relay is used as a performance metric in this paper.


{In this work, we consider the communication between a pair of source and destination terminals, where a direct link is absent, and therefore, a half-duplex relay terminal is utilized to assist the communication \cite{LTE11,Dahlmanbook2}.
To forward the signal from the source to the destination, the relay employs the non-coherent amplify-and-forward (AF) protocol \cite{CoverIT79}, which is considered as the most promising solution for current and future communication systems, since it offers a reasonable tradeoff between the benefits and implementation costs \cite{SanguinettiJSAC12}.
As a result, these relays have already been incorporated in Universal Mobile Telecommunications System (UMTS) and Release 8 of Long-Term Evolution (LTE) in the form of repeaters \cite{LTE11}.
For the AF protocol, we make the common assumption that the destination has the instantaneous channel state information (CSI) of the relay-to-destination link and the aggregate source-to-relay-to-destination link \cite{MorgenIT07,CuiGlobecom07}.
Consequently, the destination can obtain the CSI of the relay-to-destination link, which is however not available at the relay.
To secure messages against the relay, we propose a novel AN scheme, where the destination designs and broadcasts AN symbols to the relay simultaneously with the transmission of the symbols from the source.
It has been shown that a positive secrecy rate can be achievable in this AN assisted untrusted relay network \cite{HeGlobcom08}.
For this setup and with the knowledge of instantaneous CSI as assumed in Section \ref{sec:AN_SER}, we first investigate the problem of how to optimally design the AN to maximize the SER of QAM signals at the relay.
Note that the requirement of the instantaneous CSI might be strong in some practical systems.
To reduce the requirement, we then investigate the AN design based on statistical CSI in Section \ref{sec:AN_ASER}, i.e., the channel variances.
For fading channels, an important performance metric is the average SER (ASER), which quantifies the average decoding error performance over fading channels.
Interestingly, for both CSI schemes, our study shows that it is not optimal to generate AN from a Gaussian distribution.
For example, based on instantaneous CSI, we note that a QAM or a rotated QAM AN generating method maximizes the SER at the relay.
The results in this paper can be used as benchmarks for future analyses of AN-based techniques.}
The main contributions of this work can be summarized as follows:
\begin{itemize}

\item By assuming that instantaneous CSI is available at the untrusted relay and the destination, upon receiving both of the QAM symbols from the source and the AN from the destination, an exact expression for the SER at the relay when decoding the QAM symbols is first derived. Under an average power constraint, this expression is then utilized to obtain the phase and power distribution of the AN symbols to maximize the SER performance at the relay.

\item For the case where only statistical CSI is available, we first derive an exact expression for the ASER performance at the relay. Next, the optimal power distribution of the AN symbols that maximizes the ASER performance at the relay is determined.
Numerical and simulation results demonstrate that the proposed optimal AN designs guarantee significant  error rate performance enhancement compared with conventional AN designs, such as Gaussian distribution.

\end{itemize}

{Furthermore, the AN-based scheme studied in this paper is formulated according to the framework of current cellular standards such as LTE/LTE-Advanced of 4G and the next major phase of mobile telecommunications standard 5G \cite{Dahlmanbook2,YangTranMag15}. }
By applying an additional processing unit to generate AN, the studied scheme can be easily embedded in
a practical system to secure wireless communications, e.g., key transmission or control signaling.
Moreover, our study provides exact SER expressions of the QAM signals at the relay, which can be used as benchmarks for future extensions, e.g., deriving SER expressions for other modulation schemes, and designing AN signaling in other scenarios.

{\it Notation:} Throughout this paper, we use $\mathcal{CN}(b_1,b_2)$ to represent a complex circularly symmetric Gaussian distribution with mean ${b_1}$ and variance $b_2$. We use ${\mathcal E} \left\{ \cdot \right\}$ and $\Pr(\cdot)$ to denote statistical expectation and probability. Moreover, $|b|$ and $b^*$ represent absolute value and complex conjugate of $b$, respectively.

\section{System Model and Problem Formulation}
\label{sec:sm}

Let us consider a relay communication channel as shown in Figure~\ref{fig:sm}.
A legitimate transmitter (the source) sends information symbols
to a legitimate receiver (the destination) assisted by an untrusted relay.
We assume that a direct link between the legitimate terminals is not available due to, for example, high pathloss.
All the terminals are configured with a single antenna\footnote{Note that this work can be generalized to the scenario where at least one of the source and destination has multiple antennas. Whilst beyond the scope of this work, the corresponding precoding design at each terminal remains an interesting topic for future extension.}.
We assume quasi-static Rayleigh fading channels and the
channel coefficients for the source-to-relay and relay-to-destination links are respectively denoted by $h$ and $g$, where $h\sim{\cal CN}\left(0,\sigma^2_h\right)$ and $g\sim{\cal CN}\left(0,\sigma^2_g\right)$.
All channels are assumed to be reciprocal \cite{GuillaudISSP05} and
constant within the transmission duration from the source to the destination.
The transmission from the source to destination can be partitioned into two time slots.
During the first time slot, the source broadcasts the symbol $m$.
Meanwhile, to prevent the relay from \textcolor{black}{deciphering} the information, the destination sends the \textcolor{black}{AN symbols $z$ to increase the noise level} at the relay.
For this purpose, all the terminals are assumed to be perfectly synchronized. Similar to physical-layer network coding \cite{Liew2013physical}, the synchronization is an important issue, which requires further investigation.
At the end of the first time slot, the received baseband signal at the relay can be mathematically expressed as
\begin{equation}
{y}={ h} m+{ g} z+{n} \label{signalmodel}
\end{equation}
where 
the additive white Gaussian noise (AWGN) symbol $n$ is assumed to be zero mean and have the one-sided power spectral density ${N}_0$ in Watts/Hz.
During the second time slot, the source is muted, and
the relay applies the amplify-and-forward protocol \cite{CoverIT79} to forward a scaled version of its received signal to the destination.
Upon receiving this signal, the destination removes the AN $z$ prior to decoding $m$.
For the afore-mentioned communication protocol, the destination is required to have the perfect CSI of $h$ and $g$ in order to maximize the system's transmission rate \cite{LanemanIT04}.
{To be more specific, the destination performs the channel estimation of the relay-to-destination link and the compound source-to-relay-to-destination link, and the channel coefficient of the source-to-relay link is then obtained \cite{CuiGlobecom07}.}

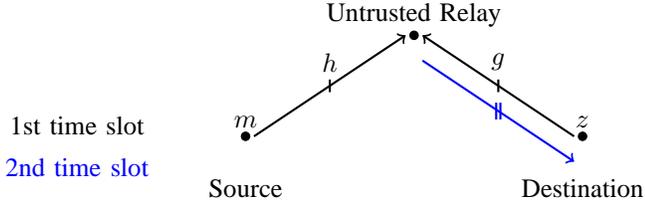
\begin{figure}
\begin{center}
{
\begin{tikzpicture}[xscale=0.56,yscale=0.56]

\draw [fill=black,black] (0, -.2) circle (0.1);
\draw [fill=black,black] (8, -.2) circle (0.1);

\node [below, black] at (0,-1) {\begin{normalsize}Source\end{normalsize}};
\node [below, black] at (4,3.2) {\begin{normalsize}Untrusted Relay\end{normalsize}};
\node [below, black] at (8,-1) {\begin{normalsize}Destination\end{normalsize}};

\node [below, black] at (0,0.5) {\begin{normalsize}$m$\end{normalsize}};
\node [below, black] at (8,0.5) {\begin{normalsize}$z$\end{normalsize}};

\draw [->,line width=0.8,black]  (0.2,-0.2)--(3.8,2.2);
\draw [fill=black,black] (4, 2.2) circle (0.1);
\draw [<-,line width=0.8,black]  (4.2,2.2)--(7.8,-0.2);


\draw [->,line width=0.8,blue]  (4.2,1.6)--(7.8,-0.8);

\draw [-,line width=0.8,black]  (2,1.15)--(2,0.85);
\draw [-,line width=0.8,black]  (6,1.15)--(6,0.85);

\draw [-,line width=0.8,blue]  (5.95,0.2)--(5.95,0.6);
\draw [-,line width=0.8,blue]  (6.05,0.2)--(6.05,0.6);

\node [below, black] at (2,2) {\begin{normalsize}$h$\end{normalsize}};
\node [below, black] at (6,2) {\begin{normalsize}$g$\end{normalsize}};

\node [below, black] at (-4,0.5) {\begin{normalsize}1st time slot\end{normalsize}};
\node [below, blue] at (-4,-0.5) {\begin{normalsize}2nd time slot\end{normalsize}};

\end{tikzpicture}
}

\end{center}

\caption{The considered system model in this work.}\label{fig:sm}
\end{figure}

While exploiting the untrusted relay to help the data transmission, the system is designed to ensure that the untrusted relay cannot decipher the source's symbol $m$.
In addition, we assume that the relay can obtain knowledge about the modulation scheme of the source, e.g., by tracking the common control channel of the network.
Moreover, it is important to note that the AN $z$ in Eq.~(\ref{signalmodel}) 
is solely designed by the destination so that the relay is not aware of the existence of the AN, neither the AN distribution.
Therefore, when decoding the symbols transmitted by the source, the relay treats the AN from the destination as noise.
Furthermore, we assume that the source
transmits demodulation reference signals (DM-RS) \cite{Dahlmanbook2} so that
the relay can perfectly estimate $h$ to perform ideal coherent
demodulation as
\begin{align}
{\tilde  y} =\frac{{h}^* }{{|{ h}|}} { y} =m+\frac{{h}^* { g} }{{|{ h}|}} z+\frac{{ h}^* }{{|{ h}|}} { n}.\label{e5}
\end{align}
Forwarded by the relay, the destination also has access to $h$ and the DM-RS of the source in order to demodulate the signal $m$.
For the source's symbol $m$, we consider square $M$-QAM modulation types with $M=4^k$ and $k=1,2,\dots$, 
which are frequently used in the current and upcoming communication standards \cite{Dahlmanbook2}.
We also make the common assumption that constellation points are uniformly distributed \cite{Proakisbook}.
In addition, we use $E_m$ and $E_z$ to denote the average energy per symbol for the source's signal $m$ and the AN $z$ sent by the destination, respectively.
Finally, $E_m$ is assumed to be known at the destination for the AN design.

Since the destination has perfect knowledge of the instantaneous channel coefficients $h$ and $g$, this knowledge can be efficiently used to design the AN symbols $z$.
The relay uses the ideal coherent demodulation with minimum distance detection to recover the source's signal $m$, and the corresponding SER is the performance metric used in this paper.
The objective is to find out the AN symbols $z$ at the destination in order to maximize the SER performance of the square $M$-QAM modulation at the relay, which we address in Section \ref{sec:AN_SER}.
To reduce the complexity of the AN design, the AN symbols $z$ can also be obtained based on long-term CSI of $h$ and $g$, i.e., $\sigma^2_h$ and $\sigma^2_g$.
In this case, the AN symbols $z$ can be determined to maximize the ASER at the relay, and this is investigated in Section \ref{sec:AN_ASER}.

\section{AN Maximizing Relay's SER Performance}
\label{sec:AN_SER}

The SER analysis in this section consists of three parts.
In Section \ref{sec:sergs}, for given instantaneous channel realizations $h$ and $g$, we derive the SER expression at the relay for a given AN symbols
$z$ of square $M$-QAM modulation.
Based on this expression, in Section \ref{sec:sersph},
we further study the problem of how to select the phase of $z$ given its amplitude $|z|$.
The problem of assigning power to the AN
symbols $z$ is considered in Section \ref{sec:AN_amplitude}, where we
derive the optimal distribution of the power of $z$, i.e., $|z|^2$, based on an average power constraint $E_z$.

\subsection{SER Expression for a Given $z$}
\label{sec:sergs}

The SER performance of the square $M$-QAM 
modulated signaling over AWGN \cite{Proakisbook} and over fading channels \cite{Simon05} has been widely studied.
To provide a comprehensive study on the SER performance at the relay, we first investigate the AN design using the instantaneous CSI in this section, and our SER analysis is based on the procedure presented in \cite[Chapter 5]{Proakisbook}.

As an example, the constellation diagram for the square $16$-QAM modulation is illustrated in Figure~\ref{fig:cd}.
Denote the minimum distance between two constellation points as $2a$ ($a>0$), then for a general $M$-QAM constellation, the average energy per signal symbol $E_m$ can be expressed as \cite[Eq.~(5-2-76)]{Proakisbook}
\begin{equation}
E_m=\frac{2}{3}a^2 T_m \left(M-1\right) \label{e6}
\end{equation}
where $T_m$ denotes the symbol duration.

\begin{figure}
\begin{center}
\resizebox{7cm}{!}{
\begin{tikzpicture}[xscale=2.5,yscale=2.5]
\draw [->,line width=0.8] (-1.1,0)--(1.4,0);
\draw [->,line width=0.8] (0,-1.1)--(0,1.4);
\node [above] at (1.3,0) {\begin{large}${\rm I}$\end{large}};
\node [right] at (0,1.3) {\begin{large}${\rm Q}$\end{large}};
\draw [fill=red,red] (-0.9487, -0.9487) circle (0.03);
\node [above] at (-0.9487, -0.9487) {\begin{small}$2$\end{small}};
\draw [fill=blue,blue] (0.3162, 0.3162) circle (0.03);
\node [above] at (0.3162, 0.3162)  {\begin{small}$13$\end{small}};
\draw [fill=blue,blue] (-0.3162, 0.3162) circle (0.03);
\node [above] at (-0.3162, 0.3162)  {\begin{small}$5$\end{small}};
\draw [fill=blue,blue] (0.3162, -0.3162) circle (0.03);
\node [above] at (0.3162, -0.3162)  {\begin{small}$15$\end{small}};
\draw [fill=blue,blue] (-0.3162, -0.3162) circle (0.03);
\node [above] at (-0.3162, -0.3162)  {\begin{small}$7$\end{small}};
\draw [fill=black] (0.9487, 0.3162) circle (0.03);
\node [above] at (0.9487, 0.3162)  {\begin{small}$9$\end{small}};
\draw [fill=black] (0.9487, -0.3162) circle (0.03);
\node [above] at (0.9487, -0.3162)  {\begin{small}$11$\end{small}};
\draw [fill=black] (-0.9487, -0.3162) circle (0.03);
\node [above] at (-0.9487, -0.3162)  {\begin{small}$3$\end{small}};
\draw [fill=black] (-0.9487, 0.3162) circle (0.03);
\node [above] at (-0.9487, 0.3162) {\begin{small}$1$\end{small}};
\draw [fill=black] (0.3162, 0.9487) circle (0.03);
\node [above] at (0.3162, 0.9487)  {\begin{small}$12$\end{small}};
\draw [fill=black](0.3162, -0.9487) circle (0.03);
\node [above] at  (0.3162, -0.9487)  {\begin{small}$14$\end{small}};
\draw [fill=black]  (-0.3162, -0.9487) circle (0.03);
\node [above] at  (-0.3162, -0.9487)  {\begin{small}$6$\end{small}};
\draw [fill=black] (-0.3162, 0.9487) circle (0.03);
\node [above] at (-0.3162, 0.9487)  {\begin{small}$4$\end{small}};
\draw [fill=red,red] (-0.9487, 0.9487) circle (0.03);
\node [above] at (-0.9487, 0.9487)  {\begin{small}$0$\end{small}};
\draw [fill=red,red] (0.9487, 0.9487) circle (0.03);
\node [above] at (0.9487, 0.9487)  {\begin{small}$8$\end{small}};
\draw [fill=red,red] (0.9487, -0.9487) circle (0.03);
\node [above] at (0.9487, -0.9487)  {\begin{small}$10$\end{small}};
\draw (0,0.9487)--(0.03,0.9487);
\node  [right] at (0.0,0.3162) {\begin{scriptsize}$a$\end{scriptsize}};
\draw (0,0.3162)--(0.03,0.3162);
\node  [right] at (0.0,0.9487) {\begin{scriptsize}$3a$\end{scriptsize}};
\draw (0,-0.3162)--(0.03,-0.3162);
\node  [right] at (0.0,-0.3162) {\begin{scriptsize}$-a$\end{scriptsize}};
\draw (0,-0.9487)--(0.03,-0.9487);
\node  [right] at (0,-0.9487) {\begin{scriptsize}$-3a$\end{scriptsize}};
\node  [below] at (0.05,0) {\begin{scriptsize}$0$\end{scriptsize}};
\draw (0.3162,0)--(0.3162,0.03);
\node  [above] at (0.3162,0) {\begin{scriptsize}$a$\end{scriptsize}};
\draw (0.9487,0)--(0.9487,0.03);
\node  [above] at (0.9487,0) {\begin{scriptsize}$3a$\end{scriptsize}};
\draw (-0.3162,0)--(-0.3162,0.03);
\node  [above] at (-0.3162,0) {\begin{scriptsize}$-a$\end{scriptsize}};
\draw (-0.9487,0)--(-0.9487,0.03);
\node  [above] at (-0.9487,0) {\begin{scriptsize}$-3a$\end{scriptsize}};
\end{tikzpicture}
}
\end{center}
\caption{The constellation diagram of square $16$-QAM with Gray mapping.} \label{fig:cd}
\end{figure}
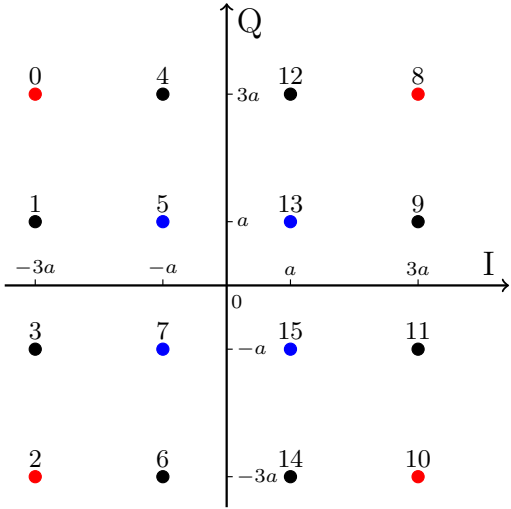

Since the AN design in this section is based on $h$ and $g$,
for simplicity of notation, we define $s:=\frac{h^*g}{|h|}z$ and $\tilde{n}:=\frac{h^*}{|h|}n$ in Eq.~(\ref{e5}),
and denote the real and imaginary parts of $s$ and $\tilde{n}$ as $s_r$, $s_i$ and $\tilde{n}_r$, $\tilde{n}_i$, respectively.
Thus, we have that $s=s_r+j s_i$ and $\tilde{n}=\tilde{n}_r+j \tilde{n}_i$ with $j=\sqrt{-1}$.
Note that the equivalent noise $\tilde{n}$ is identically distributed with $n$, and thus $\tilde{n}\sim{\cal CN}(0,2\sigma^2)$, where $\sigma^2=\frac{N_0}{2 T_m}$.
For a vertex such as the point ``0'' in Figure~\ref{fig:cd}, the received symbol $\tilde{y}$ in Eq.~(\ref{e5}) lies outside the decision region of the point ``0'' if $s_r+\tilde{n}_r\geq |h|a$ or $s_i+\tilde{n}_i\leq -|h|a$. Thus, given $s$, the error probability of the point ``0'' is given by $P_{e,0}=1-\left(1-{\Pr}\left(s_r+\tilde{n}_r\geq |h|a\right)\right)\left(1-\Pr\left(s_i+\tilde{n}_i\leq -|h|a\right)\right)$, which can be computed as
\begin{align}\label{Eq:Pe0}
P_{e,0}=1\!-\!\left(1\!-\!Q\!\left(\frac{|h|a\!-\!s_r}{\sigma}\right)\right)
\left(1\!-\!Q\left(\frac{|h|a\!+\!s_i}{\sigma}\right)\right)
\end{align}
where $Q\left(\cdot\right)$ is the Gaussian $Q$-function \cite[(4.1)]{Simon05}.
Similarly, the error probability for the $i$th constellation point $P_{e,l}$ ($l=0,1,\ldots,M-1$) is obtained by computing the probability that $\tilde{y}$ in Eq.~(\ref{e5}) lies outside its decision region.
Based on the assumption that the constellation points are uniformly distributed, for a given $z$, or equivalently for a given $s$, the SER of the square $M$-QAM signal at the relay can be derived by averaging the respective error probability expressions for all symbols under the assumption of a uniform symbol distribution, resulting in
\begin{align}
{{\rm SER}}\left(s\right)=&c\bigg[Q\left(\frac{|h|a-s_r}{\sigma}\right)+Q\left(\frac{|h|a+s_r}{\sigma}\right)\nonumber\\
&~~~+Q\left(\frac{|h|a-s_i}{\sigma}\right)+Q\left(\frac{|h|a+s_i}{\sigma}\right)\bigg]\nonumber\\
&-{c}^2\left[Q\left(\frac{|h|a-s_r}{\sigma}\right)+Q\left(\frac{|h|a+s_r}{\sigma}\right)\right]\nonumber\\
&~~~\times\left[Q\left(\frac{|h|a-s_i}{\sigma}\right)+Q\left(\frac{|h|a+s_i}{\sigma}\right)\right]\label{e23}
\end{align}
where $c=\frac{\sqrt{M}-1}{\sqrt{M}}$.
Using Eq.~(\ref{e6}), we have that
\begin{equation}
\frac{a}{\sigma}=\sqrt{\frac{3 E_m}{N_0\left(M-1\right)}}.
\end{equation}
Note that in the case of $z=0$, the SER expression in Eq.~(\ref{e23}) coincides with that of the AWGN channel \cite[Eq.~(5-2-79)]{Proakisbook}.

Before proceeding, it is interesting to note that the SER expression in Eq.~(\ref{e23}) depends on the real and imaginary parts of $s$, which motivates us to express $s$ as
\begin{align}
s={|g||z|}\exp\left(j\theta\right)\label{eq:s}
\end{align}
where we have expressed $z$, $g$ and $h$ as $z=|z|\exp(j\theta_z)$, $g=|g|\exp(j\theta_g)$, $h=|h|\exp(j\theta_h)$, respectively, and
$\theta=\theta_g-\theta_h+\theta_z$.
Now, inserting Eq.~(\ref{eq:s}) into Eq.~(\ref{e23}), the SER expression can be rewritten as a function of $\theta$ and $|z|$ as
\begin{align}
&{{\rm SER}}\left(\theta, |z|\right)=\nonumber\\
&c\bigg[\!Q\!\left(\frac{|h|a}{\sigma}
\!-\!\frac{|g||z|}{\sigma}\cos\theta \right)\!\!+\!Q\!\left(\frac{|h|a}{\sigma}\!+\!\frac{|g||z|}{\sigma}\cos\theta \right)\nonumber\\
&~~~+\! Q\!\left(\frac{|h|a}{\sigma}\!-\!\frac{|g||z|}{\sigma}\sin\theta \right)
\!\!+\!Q\!\left(\frac{|h|a}{\sigma}\!+\!\frac{|g||z|}{\sigma}\sin\theta \right)
\!\bigg]\nonumber\\
&-{c}^2\!\left[\!Q\!\left(\frac{|h|a}{\sigma}
\!-\!\frac{|g||z|}{\sigma}\cos \theta\right)\!\!+\!Q\!\left(\frac{|h|a}{\sigma}\!+\!\frac{|g||z|}{\sigma}\cos \theta\right)\!\right]\nonumber\\
&~~~\times\! \left[\!Q\!\left(\frac{|h|a}{\sigma}\!-\!\frac{|g||z|}{\sigma}\sin\theta \right)\!\!+\!Q\!\left(\frac{|h|a}{\sigma}\!+\!\frac{|g||z|}{\sigma}\sin\theta \right)\!\right].\label{e26}
\end{align}
From Eq.~(\ref{e26}), we can observe that the channel gains $|h|$ and $|g|$ play an important role on the SER performance, which we summarize in the following propositions.

{\it Proposition 1 (SER decreases in $|h|$):}
The SER in Eq.~(\ref{e26}) is a monotonically decreasing function of $|h|$.

{\it Proposition 2 (SER increases in $|g|$):}
The SER in Eq.~(\ref{e26}) is a monotonically increasing function of $|g|$.


To prove these propositions, we denote
\begin{align}
\xi(|h|,|g|)=& \sum^1_{l=0}Q\left(\frac{|h|a}{\sigma}+(-1)^l\frac{|g||z|}{\sigma}\cos\theta \right)
\end{align}
and
\begin{align}
\eta(|h|,|g|)=& \sum^1_{l=0}Q\left(\frac{|h|a}{\sigma}+(-1)^l\frac{|g||z|}{\sigma}\sin\theta \right).
\end{align}
We first focus on {\it Proposition 1}. 
Since $\xi(|h|,|g|)$ is a decreasing function of $|h|$, the first derivative of $\xi(|h|,|g|)$ with respect to $|h|$ satisfies $\frac{{\rm d}\xi(|h|,|g|)}{{\rm d}|h|}\leq 0$, and for a given $|g|$,
$\xi(|h|,|g|)\leq \xi(0,|g|)=1$.
Similarly, $\frac{{\rm d}\eta(|h|,|g|)}{{\rm d}|h|}\leq 0$ and for a given $|g|$, $\eta(|h|,|g|)\leq 1$.
In addition, since $0< c< 1$, the first derivative of the SER expression of Eq.~(\ref{e26}) with respect to $|h|$ can be written as
\begin{align}
& c\!\left(1-c\xi(|h|,|g|)\right)\frac{{\rm d}\eta(|h|,|g|)}{{\rm d}|h|}\!+\! c\!\left(1-c\eta(|h|,|g|)\right)\frac{{\rm d}\xi(|h|,|g|)}{{\rm d}|h|}\nonumber\\
& \leq  0
\end{align}
which proves {\it Proposition 1}. 
This proposition agrees with the intuition that a stronger source-to-relay link improves the decoding performance at the relay.

Regarding {\it Proposition 2}, 
for a given $|h|$, we use the first derivative of the Gaussian $Q$-function
\begin{align}
\frac{{\rm d}Q(x)}{{\rm d}x}=\frac{1}{\sqrt{2\pi}}\exp\left(-\frac{x^2}{2}\right)
 \end{align}
 to obtain
\begin{align}
\frac{{\rm d}\xi(|h|,|g|)}{{\rm d}|g|}= & \frac{\frac{|z|}{\sigma}\cos\theta}{\sqrt{2\pi}}\Bigg[\exp
\left(-\frac{1}{2}\left(\frac{|h|a}{\sigma}-\frac{|g||z|}{\sigma}\cos\theta \right)^2\right)\nonumber\\
&-\exp\left(-\frac{1}{2}\left(\frac{|h|a}{\sigma}+\frac{|g||z|}{\sigma}\cos\theta \right)^2\right)\Bigg]\nonumber\\
\geq& 0
\end{align}
which indicates that $\xi(|h|,|g|)$ is an increasing function of $|g|$.
Thus, we have
\begin{align}
\xi(|h|,|g|)\leq \lim_{|g|\to\infty}\xi(|h|,|g|)=1.
\end{align}
Therefore, since $0< c< 1$, the first derivative of the SER expression of Eq.~(\ref{e26}) with respect to $|g|$ can be derived as 
\begin{align}
&c\!\left(1-c\xi(|h|,|g|)\right)\frac{{\rm d}\eta(|h|,|g|)}{{\rm d}|g|}\!+\!c\!\left(1-c\eta(|h|,|g|)\right)\frac{{\rm d}\xi(|h|,|g|)}{{\rm d}|g|}\nonumber\\
&\geq  0
\end{align}
which proves {\it Proposition 2}. 
{This proposition suggests that a stronger relay-to-destination link deteriorates the SER performance at the relay.} 

A careful observation of Eq.~(\ref{e26}) shows that the AN design can be decoupled into the optimal design of the phase and the amplitude, respectively.
In the following subsection, we first derive the optimal rotation angle $\theta$ that maximizes the SER given by Eq.~(\ref{e26}). The value of $\theta_z$, which denotes the corresponding phase of the AN defined following Eq.~(\ref{eq:s}), is then determined accordingly.

\subsection{Selecting $\theta_z$ for a Fixed $|z|$}
\label{sec:sersph}

The focus of this subsection is to determine the optimal phase distribution of the AN symbols $z$. In this case, the power ${|z|^2}$ of the AN, is assumed to be fixed.
Then, for simplicity, we can omit $|z|$ in the argument of ${\rm SER}(\theta,|z|)$ in Eq.~(\ref{e26}) and denote the SER expression as ${\rm SER}(\theta)$ in the remainder of this subsection.

Due to the $\frac{\pi}{2}$-periodicity of the function ${ {\rm SER}}\left(\theta\right)$ and due to its symmetry, we only
consider the interval $\theta\in\left[0,\frac{\pi}{4}\right]$ to determine the maximum of ${{\rm SER}}\left(\theta\right)$.
We first numerically verify that the following result.

{\it Result 1:} ${{\rm SER}}\left(\theta\right)$ is a quasi-convex function of $\theta$
for all $\theta\in\left[0,\frac{\pi}{4}\right]$,
and either $\theta=0$ or $\theta=\frac{\pi}{4}$ maximizes the value of ${{\rm SER}}\left(\theta\right)$.

To illustrate this result, consider the constellation point ``5" in Figure~\ref{fig:cd} as an example.
Given the power of the AN symbols $z$, in order to maximize the expected SER with square $16$-QAM modulation at the relay according to {\it Result 1}, the destination can either allocate all the power ${|z|}^2$ to the direction of ``12'' or ``13'', or equally distribute this power between the directions of ``12'' and ``13''.
Due to the similarity of this AN phase design to the constellation diagram, we refer to the case of $\theta=0$ as the \emph{rotated QAM} phase selection and the latter case of $\theta=\frac{\pi}{4}$ as the \emph{QAM} phase selection. 
Accordingly, the optimal phase selection of the AN $z$ is computed as $\theta_z=\theta_h-\theta_g$ or $\theta_z=\frac{\pi}{4}+\theta_h-\theta_g$.

Next we address the question in which case using the rotated QAM constellation for the AN maximizes the SER at the relay and in which case the $M$-QAM constellation is optimal.
{Towards this aim, we obtain Figure~\ref{fig:fig2}, which shows ${|z|}/{\sigma}$ as a function of ${a}/{\sigma}$.
Each curve in this figure is plotted by numerically solving the equation ${{\rm SER}}\left(\pi/4\right)={ {\rm SER}}\left(0\right)$ for $|z|/\sigma$ for given values of $a/\sigma$.
The QAM region and the rotated QAM region denote the regions of ${{\rm SER}}\left(\pi/4\right)>{ {\rm SER}}\left(0\right)$ and ${{\rm SER}}\left(\pi/4\right)<{ {\rm SER}}\left(0\right)$, respectively.
Therefore, Figure~\ref{fig:fig2} displays phase selection thresholds for different system parameters.
In this figure, $a/\sigma$ and $|z|/\sigma$ indicate the energy of the source signal $m$ and the energy of the AN $z$, respectively.
We focus on the relation between the signal energy and the AN energy, and thus ignore the effect of channel gains by assuming $h=1$ and $g=1$.}

The curves in Figure~\ref{fig:fig2} show that for a given signal power, there exists a threshold for the AN power, under which the optimal phase selection is $\theta=0$, and above which the optimal phase selection is $\theta=\pi/4$.
To explain the intuition behind this, consider the constellation point ``5" in Figure~\ref{fig:cd}.
With the purpose of maximizing the SER performance at the relay, if the power level of the AN $|z|/\sigma$ is low (as compared to a given signal power $E_m$), the AN transmitted from the destination may deviate the equalized information signals received at the relay towards the adjacent constellation points ``1", ``4", ``13", and ``7", whilst if the power level of the AN $|z|/\sigma$ is sufficiently high, the AN may shift the receive symbol towards the farther constellation points ``0", ``12", ``3", and ``15".
From Figure~\ref{fig:fig2}, one can also observe that for small values of  $a/\sigma$, the values of the threshold are very close to zero. This fact indicates that QAM phase selection is the preferable scheme if the received signal-to-noise ratio (SNR) at the relay is low.

In particular, from Eq.~(\ref{e26}), we have
\begin{align}
\lim_{|z|\to\infty}{ {\rm SER}}\left(0, |z|\right)=& 
c+2c(1-c)Q\left(\frac{|h|a}{\sigma}\right)\nonumber\\
{\leq} &\frac{\sqrt{M}-1}{\sqrt{M}}+\frac{2(\sqrt{M}-1)}{M}\times\frac{1}{2}\nonumber\\
=&\frac{M-1}{M}\label{Eq:inftyzRotated}
\end{align}
and
\begin{align}
\lim_{|z|\to\infty}{ {\rm SER}}\left(\pi/4, |z|\right)=2c-c^2=\frac{M-1}{M}\label{Eq:inftyzQAM}
\end{align}
where the equality in (\ref{Eq:inftyzRotated}) holds only when $a=0$, i.e., when no signal is transmitted.
This indicates that when the signal is present, given sufficiently high AN noise power, the QAM phase selection yields a higher SER.
This analysis can be confirmed by Figure~\ref{fig:fig2}.
Also note that $\frac{M-1}{M}$ is the SER when the relay does not have any prior information and randomly guesses the value of signal $m$ for decoding, which serves as an upper bound on the SER of {\it any} AN scheme.
In this work, we term this as {non-informative error performance}. 
Moreover, Eqs. (\ref{Eq:inftyzRotated}) and (\ref{Eq:inftyzQAM}) reveal that when the AN power is sufficiently high, the$M$-QAM selection asymptotically achieves the non-informative error performance.

\begin{figure}
\psfrag{b}{\begin{Huge}${|z|}/{\sigma}$\end{Huge}}
\psfrag{a0}{\begin{Huge}${a}/{\sigma}$\end{Huge}}
\psfrag{Sym}{\begin{huge}QAM\end{huge}}
\psfrag{AntiSym}{\begin{huge}rotated QAM\end{huge}}
\psfrag{4QAM}{\begin{huge}$4$-QAM\end{huge}}
\psfrag{16QAM}{\begin{huge}$16$-QAM\end{huge}}
\psfrag{64QAM}{\begin{huge}$64$-QAM\end{huge}}
\resizebox{9.6cm}{!}{\includegraphics{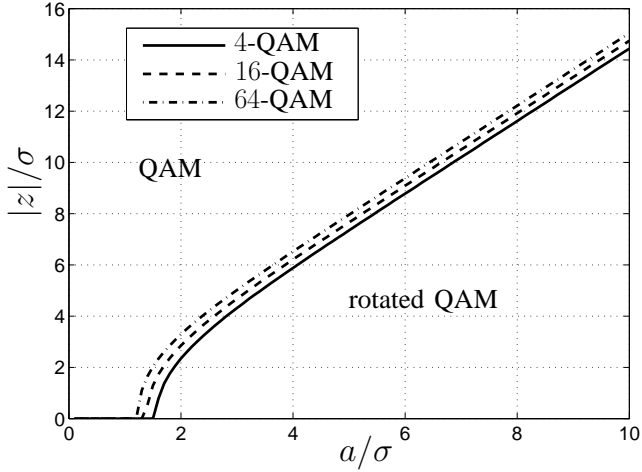}}
\caption{Thresholds for phase selection: $h=1$ and $g=1$.} \label{fig:fig2}
\end{figure}

\subsection{Assigning $|z|$ for a Given $E_z$}
\label{sec:AN_amplitude}

In Subsection \ref{sec:sersph}, we have shown how to select the phase $\theta_z$ of $z$ if the amplitude $|z|$ is a fixed value.  In general, however, it may be optimal to assign different powers to different AN symbols 
for a given average AN power. In this subsection, we assume that the average energy per AN symbol is limited to an average symbol energy ${\bar E}$, i.e.,
\begin{equation}
 E_z=T_m {\mathcal E}\left\{{|z|}^2\right\}\leq {\bar E}.\label{e28}
\end{equation}
Furthermore, we denote ${\bar P}={\bar E}/{T_m}$ as the average power of $z$.
Based on the results in Subsection \ref{sec:sersph}, the expected SER obtained from optimal phase selection is given by
\begin{equation}
 {\widetilde {\rm SER}}\left( |z|\right)=\max\left\{ {{\rm SER}}\left(\pi/4, |z|\right), { {\rm SER}}\left(0, |z|\right) \right\} \label{e29}
\end{equation}
where the ${ {\rm SER}}\left(\theta, |z|\right)$ is defined according to Eq.~(\ref{e26}).
Following the proof of {\it Proposition 2}, one can show that the first derivative of ${{\rm SER}}\left(\theta,|z|\right)$ in Eq.~(\ref{e26}) with respect to $|z|$ is non-negative, and thus ${{\rm SER}}\left(\theta,|z|\right)$ is a monotonically increasing function of $|z|$ for a given $\theta$.
Accordingly, ${\widetilde {\rm SER}}\left(|z|\right)$ in Eq.~(\ref{e29}) is also a monotonically increasing function of $|z|$.
Figure~\ref{fig:fig3} displays the values of ${\widetilde {\rm SER}}\left( |z|\right)$ for $16$-QAM and $a/\sigma=4$, where both the analytical and the simulated results are plotted.
From the figure, we can see that as $|z|/\sigma$ increases, the rotated QAM and the QAM phase selections alternatively achieve a better SER performance.
In this figure, note that $|z|^2=\bar{P}$ can be viewed as a deterministic power usage.

\begin{figure}
\centering
\psfrag{SERv}{\begin{huge}${\widetilde {\rm SER}}\left(|z|\right)$\end{huge}}
\psfrag{SER}{\begin{huge}SER\end{huge}}
\psfrag{lSER}{\begin{huge}${ {\rm SER}}\left(0, |z|\right)$\end{huge}}
\psfrag{sSER}{\begin{huge}${ {\rm SER}}\left(\frac{\pi}{4}, |z|\right)$\end{huge}}
\psfrag{siSERl}{\begin{huge}$\theta=0$ Sim.\end{huge}}
\psfrag{siSERs}{\begin{huge}$\theta=\frac{\pi}{4}$ Sim.\end{huge}}
\psfrag{b}{\begin{Huge}${|z|}/{\sigma}$\end{Huge}}
\resizebox{9.6cm}{!}{\includegraphics{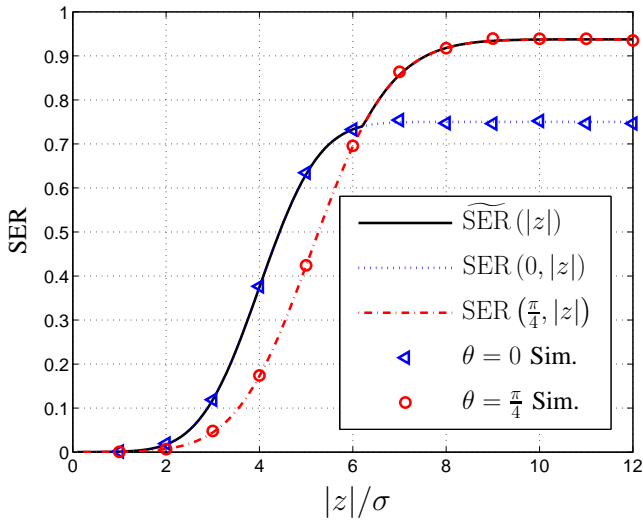}}
\caption{SER performance of $16$-QAM modulation as a function of ${|z|}/{\sigma}$, where $a/{\sigma}=4$, $h=1$, and $g=1$.} \label{fig:fig3}
\end{figure}

The objective of this subsection is to derive the optimal distribution of ${|z|}^2$ to maximize the expected value of the SER in Eq.~(\ref{e29}), given the instantaneous CSI $h$ and $g$.
Alternatively, the power needs to be smartly allocated to yield an upper bound on the SER performance with deterministic power usage. 
Denoting the probability density function (PDF) of ${|z|}^2$ as $f(\cdot)$, by taking into account the power constraint in Eq.~(\ref{e28}), the following optimization problem can be formulated
\begin{subequations}\label{P1}
\begin{align}
\max_{f\!\left(x\right)} &\int_0^{\infty}{\widetilde {\rm SER}}\left( \sqrt{x} \right) f\left(x\right) {\rm d}x \label{obj}\\
\text{s.t.} &\int_0^{\infty}x f\left(x\right){\rm d}x\leq {\bar P} \; \; {\text{(average power)}}\label{con1}\\
& \int_0^{\infty}f\!\left(x\right){\rm d}x=1 \;\; {\text{(total probability)}}\label{con2}\\
& f\left(x\right)\geq 0, \text{ for } x\geq 0 \;\; {\text{(non-negativity)}} \label{con3}
\end{align}
\end{subequations}
where the average power constraint (\ref{con1}) follows from (\ref{e28}).
For computational tractability, we only consider PDFs $f(x)$ for which all integrals in (\ref{P1}) exist.

The following theorem provides interesting insights into the power allocation problem and meanwhile can greatly simplify the computation of the PDF of $|z|$. 


{\it Theorem I (SER-maximizing power distribution):}
Any PDF $f(x)$ solving the problem (\ref{P1}) has the form
\begin{equation}
f\!\left(x\right)=\left(1-p\right) \delta\left(x-x_1\right)+p\delta\left(x-x_2\right)\label{ekey}
\end{equation}
where $\delta(x)$ is the Dirac delta function defined by \cite[Chapter 2]{Meade91}
\begin{align}
\int^\infty_{-\infty} \delta(t-\tau)g(t){\rm d}t=g(\tau)
\end{align}
for any continuous function $g$ and any value of $\tau$,
and
\begin{align}
p=\frac{\bar{P}-x_1}{x_2-x_1}\label{Eq:p}
\end{align}
with $0\leq x_1\leq {\bar P}\leq x_2$.

{\it Proof: See Appendix.} 

This theorem means that only two kinds of AN symbols are generated from the destination, one with probability $(1-p)$ and one with probability $p$, and the corresponding powers are $x_1$ and $x_2$, respectively.

Inserting Eq.~(\ref{ekey}) into Eq.~(\ref{P1}), the expected SER in Eq.~(\ref{obj}) can be computed as
\begin{align}
&{\overline {\rm SER}}(x_1,x_2)\nonumber\\
=&
\int_0^{\infty}{\widetilde {\rm SER}}\left( \sqrt{x} \right) f\left(x\right) {\rm d}x\nonumber\\
=& \, (1-p){\widetilde {\rm SER}}\left( \sqrt{x_1} \right)
+p{\widetilde {\rm SER}}\left( \sqrt{x_2} \right)
\nonumber\\
=&\, \frac{x_2-{\bar P}}{x_2-x_1}{\widetilde {\rm SER}}\left( \sqrt{x_1} \right)
+\frac{{\bar P}-x_1}{x_2-x_1}{\widetilde {\rm SER}}\left( \sqrt{x_2} \right). \label{serbar}
\end{align}

From Eq.~(\ref{serbar}), at high average AN power, i.e., $\bar{P}\to\infty$, we observe that the expected SER in Eq.~(\ref{serbar}) approaches $\frac{M-1}{M}$, which corresponds to the non-informative error performance.
The proof can be sketched as follows.
From Eqs. (\ref{Eq:inftyzRotated}) and (\ref{Eq:inftyzQAM}), we know that $\lim_{|z|\to\infty}{\widetilde {\rm SER}}\left(|z|\right)=\frac{M-1}{M}$.
Given $\bar{P}\to\infty$, by applying the {\it Optimal Power Allocation Theorem}, we have $x_2\to\infty$ due to $x_2\geq \bar{P}$.
In this case, if $x_1\to\infty$, we have $\lim_{x_1\to\infty}{\widetilde {\rm SER}}\left(\sqrt{x_1}\right)=\frac{M-1}{M}$ and $\lim_{x_2\to\infty}{\widetilde {\rm SER}}\left(\sqrt{x_2}\right)=\frac{M-1}{M}$, thereby ${\overline {\rm SER}}(x_1,x_2)$ in Eq.~(\ref{serbar}) approaches $\frac{M-1}{M}$.
If $x_1$ is a finite value, $x_2\approx \bar{P}$, then the term $\frac{x_2-{\bar P}}{x_2-x_1}{\widetilde {\rm SER}}\left( \sqrt{x_1} \right)
$ in Eq.~(\ref{ekey}) approaches zero, and the term $\frac{{\bar P}-x_1}{x_2-x_1}{\widetilde {\rm SER}}\left( \sqrt{x_2} \right)
$ in Eq.~(\ref{ekey}) converges to $\frac{M-1}{M}$, which explains the above observation.

Another interesting observation is that the maximum ${\overline {\rm SER}}(x_1,x_2)$ is a monotonically decreasing function in $|h|$.
This is because ${\overline {\rm SER}}(x_1,x_2)$ in Eq.~(\ref{serbar}) is a linear combination of ${\widetilde {\rm SER}}\left( \sqrt{x_1} \right)$
and ${\widetilde {\rm SER}}\left( \sqrt{x_2} \right)$, each of which is a monotonically decreasing function of $|h|$ as previously shown. 
Furthermore, maximizing ${\overline {\rm SER}}(x_1,x_2)$ with respect to $x_1$ and $x_2$ preserves the monotonicity.
Similarly, it can be shown that ${\overline {\rm SER}}(x_1,x_2)$ in (\ref{serbar})
is a monotonically increasing function of $|g|$.
Therefore, by applying the optimal AN design, the SER at the relay increases with the relay-to-destination link quality and decreases with the source-to-relay link quality.

The optimal values of $x_1$ and $x_2$ maximizing ${\overline {\rm SER}}(x_1,x_2)$ in Eq.~(\ref{serbar}), denoted by $x^*_1$ and $x^*_2$, can be computed numerically
based on $h$, $g$, ${\bar P}$, $M$, $a$, and $\sigma$.
Substituting $x^*_1$ and $x^*_2$ back into (\ref{serbar}) yields the maximum expected SER, which is denoted as ${\overline {\rm SER}}_{\max}$.
For example, in the case of $h=1$, $g=1$, ${\bar P}=3.9811$,
$M=16$, $a=\sqrt{10}$, and $\sigma=1/\sqrt{2}$, 
the optimal values can be computed as $x_1=0$ and $x_2=13.7098$, yielding
${\widetilde {\rm SER}}\left(\sqrt{x_1}\right)=0$ and ${\widetilde {\rm SER}}\left(\sqrt{x_2}\right)=0.5832$.
The corresponding maximum expected SER is ${\overline {\rm SER}}_{\max}=0.1694$.
Interestingly, if we use a deterministic power usage ${\bar P}$, i.e., the AN PDF can be written as $f(x)=\delta(x-{\bar P})$, 
the corresponding expected SER is $0.0371$.
Therefore, applying the {\it Theorem I} at the destination yields a SER increment of $357\%$ at the relay compared with the deterministic power usage.

{Here, we summarize the procedure to design and generate AN symbols as follows: i) 
compute the amplitudes of the AN symbols by numerically maximizing Eq.~(\ref{serbar}), ii) determine the corresponding phase value for each amplitude using Eq.~(\ref{e29}), iii) generate the AN symbols by the obtained amplitudes and phases. From the procedure, we can see that the computational complexity of the AN generation mainly stems from the optimal amplitude computation in step i). There exists numerous non-linear optimization algorithms to solve this optimization problem, such as the Newton-Raphson method, Nelder-Mead method \cite{Nelder65}, which is known as ``fminsearch'' implemented in MATLAB. For example, using ``fminsearch'' with the default termination tolerance, i.e., $10^{-4}$, the iteration number to solve the problem is generally under $100$ for the numerical examples in Section \ref{sec:sr}. Moreover, note that the computation in steps i) and ii) is required after the change of the channel coefficients $h$ or $g$.}


\subsection{Relation Between Phase Selection and Power Allocation}

Having determined the optimal phase selection and the SER-maximizing power distribution,
it is interesting to examine their interplay in the SER performance,
which we illustrate in Figures~\ref{fig:SERAN}--\ref{fig:prob_x2}.
Figure~\ref{fig:SERAN} depicts the SER performance using a deterministic power level $|z|^2=\bar{P}$ and the optimal power allocation.
The line denoted as ``Non-informative'' represents the non-informative SER performance. In the case of $16$-QAM, the non-informative SER is equal to $15/16$.
Comparing Figure~\ref{fig:fig3} and Figure~\ref{fig:SERAN}, we observe that for a given SNR $E_m/N_0$, the derived {\it Theorem I} yields an upper bound on the SER with the deterministic power usage, i.e., the curves ${{\rm SER}}\left(0, \sqrt{\bar{P}}\right)$ and ${{\rm SER}}\left(\frac{\pi}{4}, \sqrt{\bar{P}}\right)$.
To take a deeper look, we depict in Figure~\ref{fig:x1x2} the relation between the optimal power allocation ${x^*_1}$ and ${x^*_2}$.
The corresponding phase selection $\theta_{x^*_1}$ and $\theta_{x^*_2}$ are obtained from evaluating ${\widetilde {\rm SER}}\left(\sqrt{x^*_1}\right)$ and ${\widetilde {\rm SER}}\left(\sqrt{x^*_2}\right)$ in Eq.~(\ref{e29}), respectively.
Figure~\ref{fig:prob_x2} plots the probability to transmit $x^*_2$, i.e., $p$ given in Eq.~(\ref{Eq:p}), noting that the probability to transmit $x^*_1$ is given by $(1-p)$.

In correspondence to Eq.~(\ref{ekey}), we observe from Figures~\ref{fig:x1x2} and \ref{fig:prob_x2} that as the power of the AN $\bar{P}$ increases, either the probability $p$ remains constant and the powers $x^*_1$ or, $x^*_2$ increase or the values of $x^*_1$ and $x^*_2$ remain constant (with $x^*_2 > x^*_1$) and the probability $p$ of transmission with the larger power $x^*_2$ increases.
To simplify the discussion of Figures~\ref{fig:SERAN}--\ref{fig:prob_x2}, we partition the values of the AN-to-natural noise ratio (ANR) per symbol into multiple regions by introducing the transition points ${A}$--${D}$. From these figures, we obtain the following observations.

{\it Region I ($0 \leq {E_z}/{N_0} < A$):}
In this region we observe that when the ANR is low, the optimal AN design is to transmit a noise symbol with power $x^*_2$ and rotated QAM phase $\theta = 0$ at a constant probability $p_0$, and to transmit no AN ($x^*_1 = 0$) with probability $(1-p_0)$. Thus, as $\bar{P}$ increases, the power $x^*_2$ increases while the probability of transmitting AN remains constant.

{\it Region II ($A \leq {E_z}/{N_0} < B$):} When the ANR is medium, the power to transmit rotated QAM AN symbols $x^*_2$ reaches a constant, and the probability to transmit the AN symbols increases as $E_z/N_0$ increases. The corresponding SER in Figure~\ref{fig:SERAN} is higher than those using the deterministic power $\bar{P}$.

{\it Region III ($B \leq {E_z}/{N_0} < C$):} In this region, the probability to transmit AN symbols with the rotated QAM phase selection reaches $1$, and the corresponding transmitting power reaches the highest $x^*_2=\bar{P}$. 
In other words, for the AN symbols, the deterministic power usage and the rotated QAM phase selection achieve the maximum SER, which can also be observed in Figure~\ref{fig:SERAN}.

{\it Region IV ($C \leq {E_z}/{N_0} < D$):} The ANR arrives at a threshold $ {E_z}/{N_0}= C$, above which the optimal AN design is to use both of the two kinds of AN symbols: the rotated QAM phase selection with power $x^*_1$ and the QAM phase selection with power $x^*_2$.
The probability to transmit the QAM phase selection symbols increases as the ANR increases.

{\it Region V (${E_z}/{N_0} \geq D$):} When the ANR is large, the QAM phase selection with a deterministic power $\bar{P}$ achieves the maximum SER.
In particular, when ${E_z}/{N_0}$ is sufficiently large, i.e., $E_z/N_0\geq 17$ dB in Figure~\ref{fig:SERAN}, the SER converges to the non-informative SER, which was mathematically proved in Eq.~(\ref{Eq:inftyzQAM}).

To intuitively understand the above observations, again take the constellation point ``5'' in Figure~\ref{fig:cd} as an example.
When the AN power is low, the optimal AN generation scheme is to burst the limited power to move ``5'' towards the adjacent points ``1'', ``4'', ``13'' and ``7'' in order to introduce decoding errors.
As the AN power goes sufficiently large, some of the power can be used to move ``5'' towards the points ``0'', ``12'', ``3'' and ``15'' in order to induce more decoding errors.
Until when the AN power is significantly large, beaming all the power to the directions of points ``0'', ``12'', ``3'' and ``15'' yields the maximum SER.

\begin{figure}
\psfrag{A}{\begin{huge}\textcolor{blue}{$A$}\end{huge}}
\psfrag{B}{\begin{huge}\textcolor{blue}{$B$}\end{huge}}
\psfrag{C}{\begin{huge}\textcolor{blue}{$C$}\end{huge}}
\psfrag{D}{\begin{huge}\textcolor{blue}{$D$}\end{huge}}
\psfrag{SER}{\begin{Huge}SER\end{Huge}}
\psfrag{theta0}{\begin{Huge}${{\rm SER}}\left(0, \sqrt{\bar{P}}\right)$\end{Huge}}
\psfrag{thetapi4}{\begin{Huge}${{\rm SER}}\left(\frac{\pi}{4}, \sqrt{\bar{P}}\right)$\end{Huge}}
\psfrag{optimal}{\begin{Huge}${\overline {\rm SER}}_{\max}$\end{Huge}}
\psfrag{Non-informative11111111}{\begin{Huge}Non-informative\end{Huge}}
\psfrag{AN}{\begin{Huge}$\frac{E_z}{N_0}$ in dB\end{Huge}}
\resizebox{9.3cm}{!}{\includegraphics{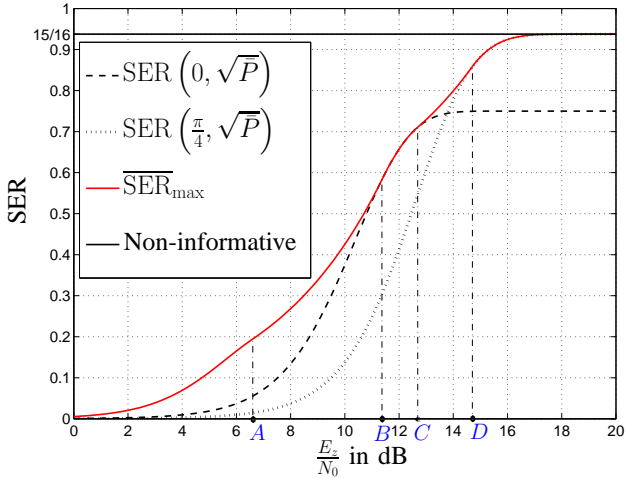}}
\caption{SER performance of $16$-QAM modulation as a function of ${E_z}/{N_0}$ without and with optimal $|z|$, where ${E_m}/{N_0}=20$ dB,
$h=1$, and $g=1$.} \label{fig:SERAN}
\end{figure}

\begin{figure}
\psfrag{A}{\begin{huge}\textcolor{blue}{$A$}\end{huge}}
\psfrag{B}{\begin{huge}\textcolor{blue}{$B$}\end{huge}}
\psfrag{C}{\begin{huge}\textcolor{blue}{$C$}\end{huge}}
\psfrag{D}{\begin{huge}\textcolor{blue}{$D$}\end{huge}}
\psfrag{theta}{\begin{Huge}$\theta$\end{Huge}}
\psfrag{x1}{\begin{Huge}$\sqrt{x^*_1}$\end{Huge}}
\psfrag{x2}{\begin{Huge}$\sqrt{x^*_2}$\end{Huge}}
\psfrag{t1}{\begin{Huge}$\theta_{x^*_1}$\end{Huge}}
\psfrag{t2}{\begin{Huge}$\theta_{x^*_2}$\end{Huge}}
\psfrag{pi/4}{\begin{Huge}\textcolor{blue}{$\frac{\pi}{4}$}\end{Huge}}
\psfrag{barP}{\begin{Huge}$\sqrt{\bar{P}}$\end{Huge}}
\psfrag{a}{\begin{Huge}$a$\end{Huge}}
\psfrag{AN}{\begin{Huge}$\frac{E_z}{N_0}$ in dB\end{Huge}}
\psfrag{Optimal |z|}{\begin{Huge}Amplitude\end{Huge}}
\resizebox{9.3cm}{!}{\includegraphics{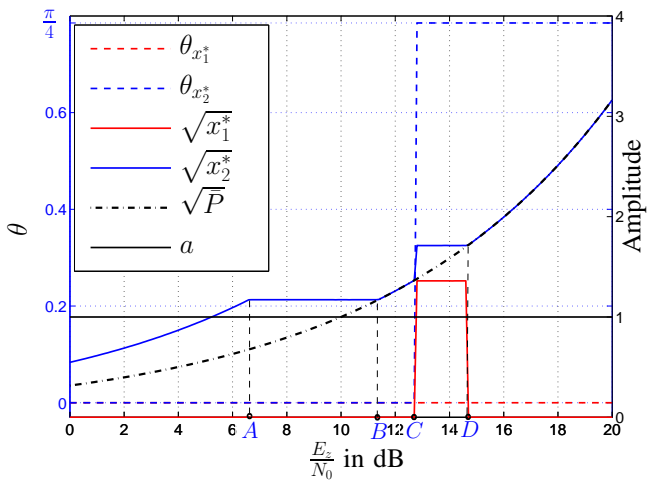}}
\caption{Optimal amplitude distribution and the corresponding phase selection for $16$-QAM modulation, where ${E_m}/{N_0}=20$ dB,
$h=1$, and $g=1$.} \label{fig:x1x2}
\end{figure}

\begin{figure}
\psfrag{A}{\begin{huge}\textcolor{blue}{$A$}\end{huge}}
\psfrag{B}{\begin{huge}\textcolor{blue}{$B$}\end{huge}}
\psfrag{C}{\begin{huge}\textcolor{blue}{$C$}\end{huge}}
\psfrag{D}{\begin{huge}\textcolor{blue}{$D$}\end{huge}}
\psfrag{p01}{\begin{Huge}\textcolor{blue}{$p_0$}\end{Huge}}
\psfrag{p2}{\begin{Huge}$p$\end{Huge}}
\psfrag{AN}{\begin{Huge}$\frac{E_z}{N_0}$ in dB\end{Huge}}
\resizebox{9.3cm}{!}{\includegraphics{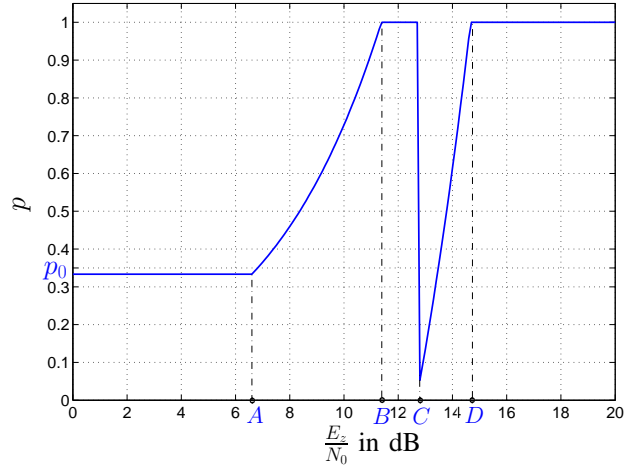}}
\caption{Probability to transmit $x^*_2$ for $16$-QAM modulation, ${E_m}/{N_0}=20$ dB,
$h=1$, and $g=1$.} \label{fig:prob_x2}
\end{figure}


\section{AN Maximizing Relay's ASER Performance}\label{sec:AN_ASER}

In Section \ref{sec:AN_SER}, we have obtained the optimal phase and amplitude distribution of the AN based on the instantaneous channel knowledge of the source-to-relay and relay-to-destination links. In this section, we consider the optimal distribution of the AN symbol power$|z|^2$ that maximizes the ASER performance of the $M$-QAM modulation at the relay provided that the long-term statistical CSI is known to the destination. Following a similar approach to Section \ref{sec:AN_SER}, we first present an analytical expression for the aforementioned performance metric over Rayleigh fading channels. The optimal phase and power distributions of the AN that maximizes this performance are then determined.

%
%


\subsection{ASER Expression for a Given $z$}

In this section, 
the channel coefficients $h$ and $g$ are assumed to be random variables.
Moreover, the envelope of the relay-to-destination channel coefficient $g$ is assumed to be Rayleigh distributed.
For a given AN symbols $z$, the received AN at the relay $gz$ can be easily shown to be distributed as $gz\sim{\cal CN}\left(0,\sigma^2_g|z|^2\right)$.
This indicates that the received AN at the relay is an extra source of the AWGN $n$.
Hence, as can be observed from the signal model in Eq.~(\ref{signalmodel}), 
the overall received noise at the relay is a superposition of the AN and the natural AWGN.
The relay-to-destination channel coefficient $g$ and the AWGN symbols $n$ are independent, and it can be easily shown that the received noise $gz+n$ is 
distributed as $gz+n\sim{\cal CN}\left(0,\sigma^2_g|z|^2+2\sigma^2\right)$, where $2\sigma^2$ is the variance of the AWGN $n$. 
For Rayleigh faded source-to-relay channel gain $|h|$ and the overall noise symbol $gz+n$, the performance of the $M$-QAM modulation at the relay can be obtained using \cite[Eq.~(8.106)]{Simon05} as 

\begin{align}\label{Eq:ASER}
{\rm ASER }(z)=&2c\left(1-\sqrt{\frac{1.5\bar{\gamma}_s(z)}{M-1+1.5\bar{\gamma}_s(z)}}\right)\nonumber\\
&-c^2\Bigg[1-\sqrt{\frac{1.5\bar{\gamma}_s(z)}{M-1+1.5\bar{\gamma}_s(z)}}\nonumber\\
&\times\left(\frac{4}{\pi}\tan^{-1}\sqrt{\frac{M-1+1.5\bar{\gamma}_s(z)}{1.5\bar{\gamma}_s(z)}}\right)\Bigg]
\end{align}
where $\bar{\gamma}_s(z)=\frac{\sigma^2_h E_m}{\sigma^2_g|z|^2+2\sigma^2}$ represents the average received SNR per symbol.

Similar to {\it Propositions 1} and {\it 2} made on Eq.~(\ref{e26}), the effects of channel statistics on the ASER given by Eq.~(\ref{Eq:ASER}) can be summarized as follows.

{\it Proposition 3 (ASER monotonicity):}
The ASER in Eq.~(\ref{Eq:ASER}) is a monotonically increasing function of $\sigma^2_g$, as well as a monotonically decreasing function of $\sigma^2_h$.

This proposition is intuitive and is similar to {\it Propositions 1} and {\it 2} 
in Section \ref{sec:AN_SER}: Stronger relay-to-destination channels help to deteriorate the ASER performance at the relay, whereas stronger source-to-relay channels improve the average decoding performance at the relay.

{\it Proposition 4 (high signal and AN power performance):}
At high SNR and high ANR, i.e., $E_m\gg \sigma^2$ and $|z|^2\gg \sigma^2$, $\bar{\gamma}_s(z)\approx \frac{\sigma^2_h}{\sigma^2_g} \frac{E_m}{|z|^2}$.

{\it Proposition 4} reveals that provided an adequately high signal and AN power is available, the {\it relative} strength of the source-to-relay and relay-to-destination links affects the ASER performance of the QAM signals at the relay.

{\it Proposition 5 (high AN power performance):}
In the high AN power regime, i.e., for $|z|^2\to\infty$, $\bar{\gamma}_s\to 0$, we have $\lim_{|z|\to\infty}{\rm ASER}(z)=2c-c^2=\frac{M-1}{M}$.

This proposition shows that given sufficiently high AN power, the ASER at the relay is close to the non-informative error performance.

Furthermore, we observe that the ASER expression in Eq.~(\ref{Eq:ASER}) is independent of the phase of the AN, which is different from the AN design in Section \ref{sec:AN_SER} in the case of instantaneous CSI.
Following a similar approach as in Section \ref{sec:AN_amplitude},
our objective in the following subsection is to derive the SER-maximizing power distribution for the AN under an average power constraint.

\subsection{Assigning $|z|$ for a Given $E_z$}

We rewrite the function ${\rm ASER}(z)$ in Eq.~ (\ref{Eq:ASER}) as ${\rm ASER}(\sqrt{x})$ to reflect the effect of AN power, where $x=|z|^2$.
Similar to Subsection \ref{sec:AN_amplitude}, the optimal power assignment problem is given by replacing $\widetilde{\rm SER}(\sqrt{x})$ in Eq.~(\ref{P1}) by ${\rm ASER}(\sqrt{x})$, and now $f(x)$ represents the PDF of $x$ for Rayleigh fading channels.
Since one can see that ${\rm ASER}(\sqrt{x})$ in Eq.~(\ref{Eq:ASER}) is a monotonically increasing function of $x$, the distribution of the optimal AN directly follows from the PDF expression derived in Eq.~(\ref{ekey}).
Therefore, substituting Eq.~(\ref{ekey}) into Eq.~(\ref{Eq:ASER}) yields the expected ASER
\begin{align}\label{Eq:AASER}
&\overline{\mathrm{ASER}}(x_1,x_2)\nonumber\\
=&\frac{x_2-\bar{P}}{x_2-x_1}{\rm ASER}(\sqrt{x_1})+\frac{\bar{P}-x_1}{x_2-x_1}{\rm ASER}(\sqrt{x_2})
\end{align}
where $\bar{P}=\frac{E_z}{T_0}$ is the average power constraint on the AN, and $0\leq x_1\leq \bar{P}\leq x_2$.
The optimal values of $x_1$ and $x_2$ maximizing $\overline{\mathrm{ASER}}(x_1,x_2)$ in Eq.~(\ref{Eq:AASER}) are then derived numerically based on ${\bar P}$, $M$, $E_m$, $\sigma$, $\sigma^2_h$, and $\sigma^2_g$, where $\sigma^2_h$ and $\sigma^2_g$ represent the statistical/long-term CSI. Note that the computational complexity of the derivation has been investigated in Section III.C.


Similar to the analysis following Eq.~(\ref{serbar}), we can also show that as the AN power increases, the expected ASER in Eq.~(\ref{Eq:AASER}) converges to the non-informative SER performance.

\section{Numerical and Simulation Results}
\label{sec:sr}

In this section, we present simulation results to validate the obtained analytical results derived in the previous sections.
In all figures, the average SNR is chosen as  $E_m/N_0=10$ dB.

In order to depict the SER performance at the relay using the optimal AN design in Section \ref{sec:AN_SER}, we plot Figure~\ref{fig:SER_M4}, where the signal symbol $m$ is randomly selected from the square $4$-QAM constellation.
We observe that all analytical results coincide very well with the corresponding simulation results.
For the sake of comparison, the expected SER
values for Gaussian AN are also plotted in Figure~\ref{fig:SER_M4}.
In the Gaussian case, we assume that the AN symbols $z$ is generated according
to a complex Gaussian distribution with the same average symbol energy $E_z$.
The line denoted as ``Non-informative'' represents the non-informative SER performance. 
The curve ``w/o AN'' in the figure depicts the SERs at the relay without AN, which lies a value of $0.0016$ for the given parameters.
Figure~\ref{fig:SER_M4} shows that the SER at the relay can be significantly increased by applying our AN design.
Even when the ANR $E_z/N_0$ is small, e.g., ${E_z}/{N_0}=2$ dB, the SER is increased from $0.0016$ to $0.05$, i.e., an increment of $303\%$.
Moreover, Figure~\ref{fig:SER_M4} clearly demonstrates that the Gaussian distribution is not optimal for our AN symbol generation and our scheme described in Section \ref{sec:AN_SER} can induce much larger SER at the relay.
For the case shown in Figure~\ref{fig:SER_M4}, if the ANR $E_z/N_0$ is above $16$ dB, the maximum SER, i.e., ${\overline {\rm SER}}_{\max}$, achieves $3/4$.
This fact indicates that if the AN $z$ is properly generated and $E_z$ is above a certain threshold, applying ideal coherent demodulation at the relay does not lead to better performance than the non-informative case, and secure data transmission can therefore be ensured by the proposed AN design.
Figure~\ref{fig:SER_M16} shows the SER performance for square $16$-QAM constellations.
Comparing Figure~\ref{fig:SER_M16} with Figure~\ref{fig:SER_M4}, similar observations can be made as for the case of $4$-QAM.

\begin{figure}
\centering
\psfrag{SERm}{\begin{Large}${\overline {\rm SER}}_{\max}$ Ana.\end{Large}}
\psfrag{SERms}{\begin{Large}${\overline {\rm SER}}_{\max}$ Sim.\end{Large}}

\psfrag{SERg}{\begin{Large}Gaussian Ana.\end{Large}}
\psfrag{SERgs}{\begin{Large}Gaussian Sim.\end{Large}}

\psfrag{Random Guess}{\begin{Large}Non-informative\end{Large}}
\psfrag{SER wo AN}{\begin{Large}w/o AN\end{Large}}


\psfrag{SER}{\begin{huge}${\rm SER}$\end{huge}}
\psfrag{AN}{\begin{huge}$\frac{E_z}{N_0}$ in dB\end{huge}}
\resizebox{9.6cm}{!}{\includegraphics{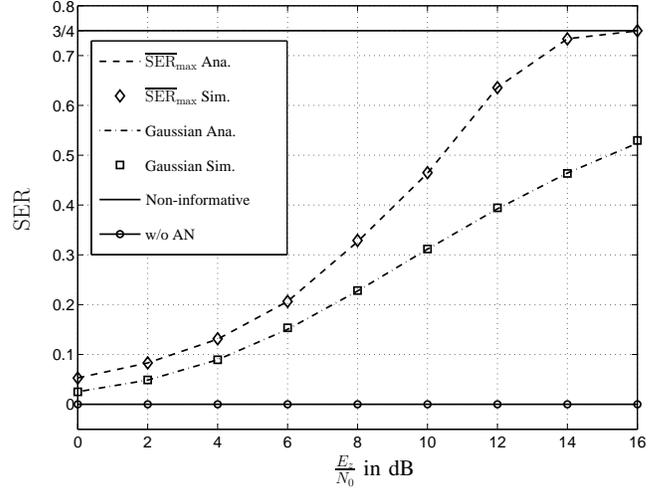}}
\caption{SER performance of $4$-QAM modulation as a function of ${E_z}/{N_0}$, where $E_m/N_0=10$ dB,
$h=1$, and $g=1$.} \label{fig:SER_M4}
\end{figure}

\begin{figure}
\centering
\psfrag{SERg}{\begin{Large}Gaussian Ana.\end{Large}}
\psfrag{SERgs}{\begin{Large}Gaussian Sim.\end{Large}}
\psfrag{Random Guess}{\begin{Large}Non-informative\end{Large}}
\psfrag{SERms}{\begin{Large}${\overline {\rm SER}}_{\max}$ Sim.\end{Large}}
\psfrag{SERm}{\begin{Large}${\overline {\rm SER}}_{\max}$ Ana.\end{Large}}
\psfrag{SERn}{\begin{Large}w/o AN\end{Large}}
\psfrag{SER}{\begin{huge}${\rm SER}$\end{huge}}
\psfrag{AN}{\begin{huge}$\frac{E_z}{N_0}$ in dB\end{huge}}


\resizebox{9.6cm}{!}{\includegraphics{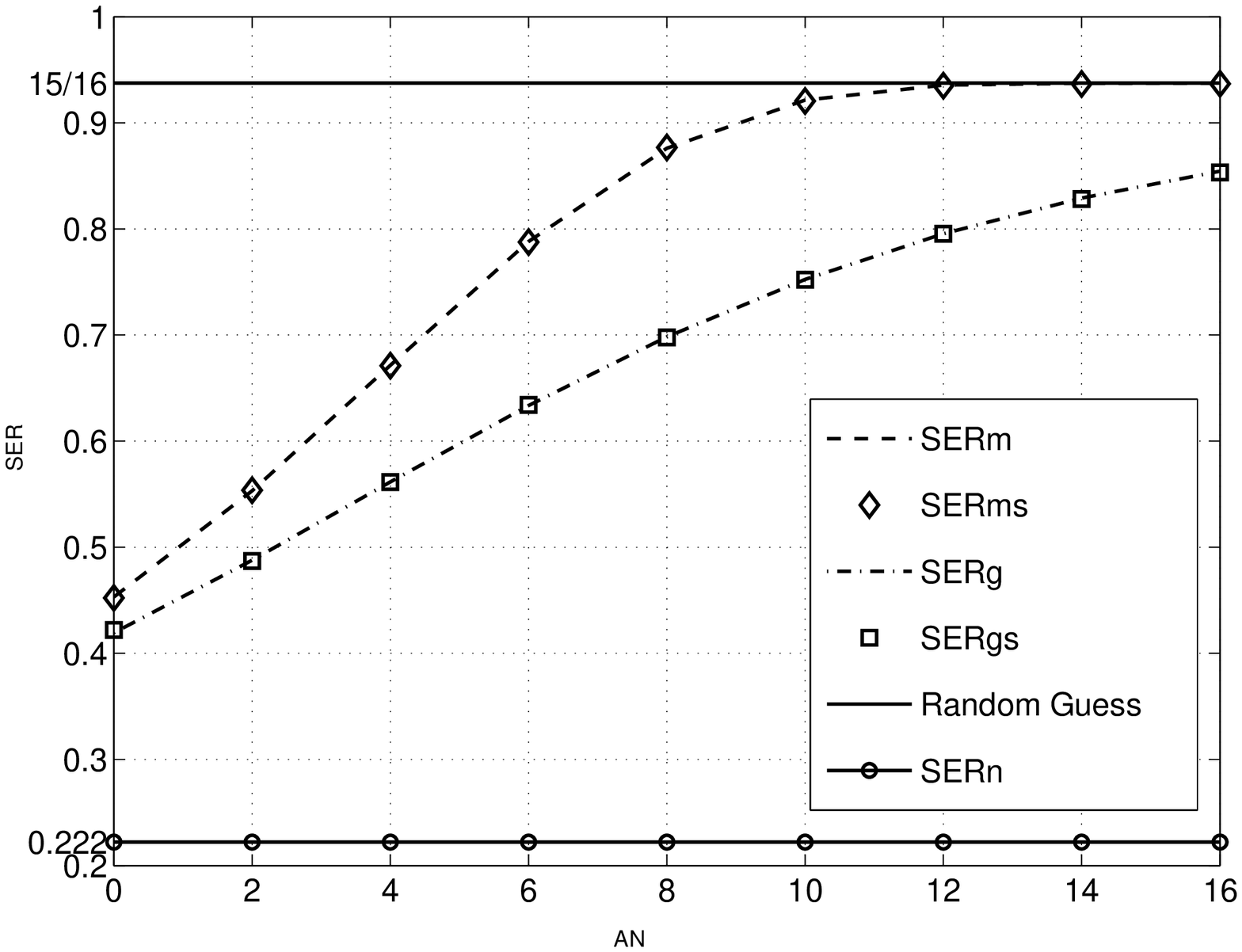}}
\caption{SER performance of $16$-QAM modulation as a function of ${E_z}/{N_0}$, where $E_m/N_0=10$ dB,
 $h=1$, and $g=1$.} \label{fig:SER_M16}
\end{figure}

\begin{figure}
\centering
\psfrag{SERm}{\begin{Large}${\overline {\rm ASER}}_{\max}$ Ana.\end{Large}}
\psfrag{SERms}{\begin{Large}${\overline {\rm ASER}}_{\max}$ Sim.\end{Large}}

\psfrag{SERu}{\begin{Large}Uniform Num.\end{Large}}
\psfrag{SERus}{\begin{Large}Uniform Sim.\end{Large}}

\psfrag{SERe}{\begin{Large}Exponential Num.\end{Large}}
\psfrag{SERes}{\begin{Large}Exponential Sim.\end{Large}}

\psfrag{Random Guess}{\begin{Large}Non-informative\end{Large}}

\psfrag{M=16}{\begin{Large}$M=16$\end{Large}}
\psfrag{M=4}{\begin{Large}$M=4$\end{Large}}

\psfrag{SER}{\begin{huge}${\rm ASER}$\end{huge}}
\psfrag{AN}{\begin{huge}$\frac{E_z}{N_0}$ in dB\end{huge}}

\resizebox{9.6cm}{!}{\includegraphics{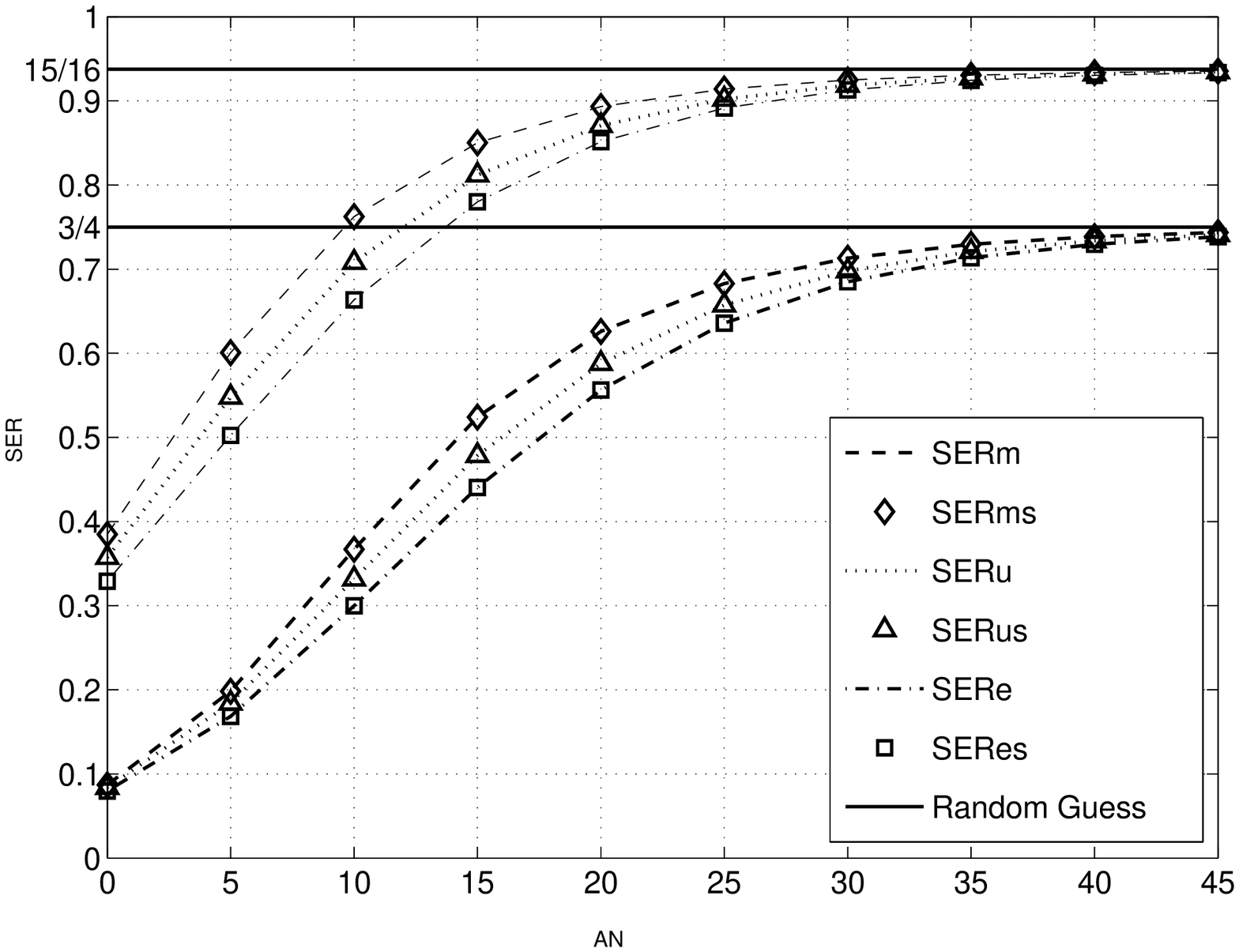}}
\caption{ASER performance of $4$-QAM and $16$-QAM modulations as a function of ${E_z}/{N_0}$, where $E_m/N_0=10$ dB,
$\sigma_h=1$, and $\sigma_g=1$.} \label{fig:ASER}
\end{figure}

\begin{figure}
\centering
\psfrag{SERm}{\begin{Large}Instantaneous CSI Based Sim.\end{Large}}

\psfrag{SERg}{\begin{Large}Gaussian Sim.\end{Large}}

\psfrag{SERfm}{\begin{Large}Statistical CSI Based Ana.
\end{Large}}
\psfrag{SERfg}{\begin{Large}Exponential Num.\end{Large}}

\psfrag{Random Guess}{\begin{Large}Non-informative\end{Large}}

\psfrag{SER}{\begin{huge}${\rm ASER}$\end{huge}}

\psfrag{AN}{\begin{huge}$\frac{E_z}{N_0}$ in dB\end{huge}}

\resizebox{9.6cm}{!}{\includegraphics{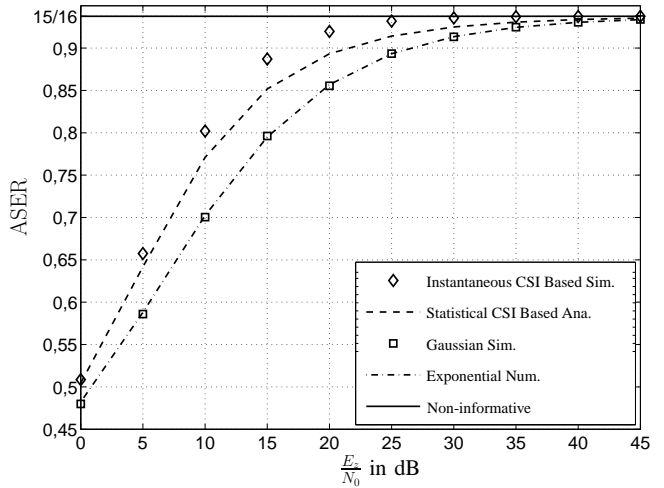}}
\caption{ASER performance of $16$-QAM modulation as a function of $E_z/N_0$, where $E_m/N_0=10$ dB,
$\sigma_h=1$, and $\sigma_g=1$.} \label{fig:ASER_compare}
\end{figure}

Now, we investigate the ASER performance at the relay using the optimal AN design in Section \ref{sec:AN_ASER}.
Figure~\ref{fig:ASER} depicts the ASER performance for square $4$-QAM and $16$-QAM signals.
In this figure, $\overline{\rm ASER}_{\max}$ is computed by numerically maximizing $\overline{\mathrm{ASER}}(x_1,x_2)$ in (\ref{Eq:AASER}) with respect to $x_1$ and $x_2$.
{Before comparing with other power distributions, we note that from numerical results, the optimal power allocation for the considered scenarios in Figure~\ref{fig:ASER} is $x^*_2=\bar{P}$ with probability $p=1$.
In other words, different from the AN design based on the instantaneous CSI in Section \ref{sec:AN_SER}, a constant power usage is optimal to design the AN based on the statistical CSI.
Intuitively, due to the averaging effect over several channel realizations, the ASER performance is not sensitive to the different instantaneous power allocations.}
For comparison, the ASER curves for various power distributions are plotted, i.e.,
the uniform and the exponential power distribution.
For each curve, the numerical results are obtained from computing $\int^\infty_0{\rm ASER}(\sqrt{x})f(x){\rm d}x$, where
\begin{align}
f(x)=\frac{1}{2\bar{P}}, \ \text{ for } 0\leq x\leq 2\bar{P}
\end{align}
and
\begin{align}
f(x)=\frac{1}{\bar{P}}\exp\left(-\frac{x}{\bar{P}}\right),\ \text{ for } x\geq 0
\end{align}
are the PDFs of the AN power with uniform and exponential distributions, respectively.
Note that there exists numerous AN designs with exponentially distributed power, such as the Gaussian distributed AN.
Figure~\ref{fig:ASER} shows that our proposed scheme with the SER-maximizing power distribution yields the largest ASER at the relay.
Given sufficiently high ANR, all the ASER curves achieve the non-informative error performance, which confirms {\it Proposition 5}. 
However, 
the performance difference between different AN designs is not significant.
This can be explained by the fact that the ASER expression in (\ref{Eq:ASER}) does not consider the phase of the AN.
Therefore, the degree of freedom of the phase design for the AN is not utilized.

Furthermore, it is interesting to compare the SER performance of the optimal AN designs in Sections \ref{sec:AN_SER} and \ref{sec:AN_ASER}.
Figure~\ref{fig:ASER_compare} compares the ASER performance of various AN designs. In the figure, ``Instantaneous CSI Based Sim.'' curve is plotted by computing $\mathcal{E}_{h,g}[{\overline{\rm SER}}_{\max}]$ using the Monte Carlo method,
and the curve ``Statistical CSI Based Ana.'' 
is plotted using the same method as in Figure~\ref{fig:ASER}.
The curve ``Gaussian Sim.'' is plotted by numerically computing $\mathcal{E}_{h,g}[{{\rm SER}}(z)]$, with Gaussian distributed AN.
From Figure~\ref{fig:ASER_compare}, we observe that the instantaneous CSI based AN design yields a higher ASER than the statistical CSI based AN design, which is consistent with our intuition.
However, the performance difference between these two designs is not quite significant.
Without optimal phase and power designs, the Gaussian distribution performs worse than both CSI based AN designs.

\section{Conclusions}
\label{sec:cs}

In this paper, we have investigated physical layer secrecy for a two-hop
single-antenna relay channel, where one source aims to transmit to one destination assisted by one untrusted relay.
{In order to achieve physical layer secrecy, a novel AN scheme was proposed to enhance the secrecy of wireless communications from the source to the destination.}
The objective was to design AN symbols generated by the destination node that degrades the error probability performance at the untrusted relay, in particular, that maximizes its achieved SER performance.
For the case where perfect
instantaneous CSI of the source-to-relay and relay-to-destination links is available
at the destination, we have derived exact analytical SER expressions for the relay
and studied the optimal design of the AN signal to maximize the corresponding SER.
In this work, the source adopts a squared $M$-QAM modulation scheme as in current
modern cellular standards. It was shown that the Gaussian distribution, which is
frequently used in the context of AN \cite{Goel,ZhouTVT10,ZhangTWC15}, is not optimal in general. The
optimal AN for the considered relay channel was found to be QAM or rotated QAM phase
selection. Moreover, compared with the Gaussian AN, our optimal AN distribution can
yield remarkably higher SER at the relay. For the case where the AN design is based
on the long-term CSI of the source-to-relay and relay-to-destination links, we
presented the corresponding ASER expression at the relay, and the optimal AN
distribution to maximize the ASER performance was determined accordingly.
Interestingly, it was shown that the phase of the AN does not affect the ASER
performance at the relay. Regarding the design of the power distribution of the AN,
the obtained power distribution was shown to deliver improved ASER performance than
various other power distributions, such as the uniform and exponential
distributions. Finally, we mention that there are many directions for further
extensions of this work. For example, when perfect CSI of the source-to-relay and
relay-to-destination links is not available, opportunistic scheduling with low-rate
CSI feedback might be applied  \cite{LiTComm10}. Other interesting directions include the AN
design for imperfect CSI knowledge as well as for the multiple-antenna relay channel
and two-way communication networks.

\section*{Appendix: Proof of {Theorem I}}

In the following, we base our proof on the concept of Linear Programming and the Karush-Kuhn-Tucker (KKT) optimality conditions \cite[p. 243]{Boyd}.

Before proceeding, note that for a discrete realization $x_1$ with probability $p$, one can express $f(x)$ at $x_1$ as
\begin{equation}
 f(x)=p\delta\left(x-x_1\right).
\end{equation}
If a PDF $f(x)$, which solves the problem (\ref{P1}), has the form of $f(x)\!=\!\delta\left(x-{x_0}\right)$, we can easily obtain $x_0\!=\!{\bar P}$, as ${\widetilde {\rm SER}}\left(x\right)$ in (\ref{e29}) is a monotonically increasing function.
This case is trivially contained in {\it Theorem I}.
In the following, we will assume that $f(x)>0$ for at least two different values of $x$. In this case, at the optimum, the constraint (\ref{con1})  must be met with equality as otherwise we can further increase the objective function (\ref{obj}) by
increasing $f\!\left(x_1\right)$ and decreasing $f\!\left(x_2\right)$ for some $x_1>x_2$ without violating any constraint, which contradicts to the optimality assumption.

If the optimal distribution $f(x)$ corresponds to that of a discrete random variable with $n$ realizations, $f(x)$ can be expressed as
\begin{equation}
f\!\left(x\right)=\sum_{i=1}^n p_i\delta\left(x-x_i\right)  \label{e35}
\end{equation}
where $\sum_{i=1}^{n} p_i=1$ and $p_i>0$. Inserting (\ref{e35}) into (\ref{obj})-(\ref{con2}), we can obtain the following system of linear equations
\begin{subequations}\label{P2}
\begin{align}
\alpha_1 p_1 +\alpha_2 p_2+\cdots +\alpha_n p_n&=b \\
 \beta_1 p_1 +\beta_2 p_2+\cdots  +\beta_n p_n &={\bar P}\label{esys2a}\\
  p_1 + p_2+\cdots  + p_n& =1 \label{esys2}
\end{align}
\end{subequations}
where $\alpha_i={\widetilde {\rm SER}}\left(\sqrt{x_i}\right)$, $\beta_i=x_i$, and $0\leq p_i\leq 1$
with $i=1,\ldots,n$. The achieved maximum value of the objective function (\ref{obj}) is denoted as $b$.

If the optimal PDF $f(x)$ contains at least one interval $\left[x_a, x_b\right]$ with $f(x)>0$ for any $x\in\left[x_a, x_b\right]$, we can divide $\left[x_a, x_b\right]$ into $m$ non-overlapping sub-intervals
$\left({\hat x}_{m-1},{\hat x}_{m}\right)$ with $x_a={\hat x}_{0}<{\hat x}_{1}<\cdots<{\hat x}_{m}=x_b$. Then, following the first mean value theorem \cite[Theorem 12.111]{Grad}, we can find some ${\tilde x}_i$ and ${\bar x}_i$ with ${\hat x}_{i-1}\leq {\tilde x}_i\leq {\hat x}_{i}$ and
${\hat x}_{i-1}\leq {\bar x}_i\leq {\hat x}_{i}$ such that
\begin{equation}
 p_i=\int_{{\hat x}_{i-1}}^{{\hat x}_{i}} f(x)\,{\rm d}x
\end{equation}
\begin{equation}
\int_{x_a}^{x_b}{\widetilde {\rm SER}}\left( \sqrt{x} \right) f\left(x\right)\, {\rm d}x=\sum_{i=1}^m {\widetilde {\rm SER}}\left(\sqrt{{\tilde x}_i}\right) p_i
\end{equation}
and
\begin{equation}
\int_{x_a}^{x_b}x f\!\left(x\right)\,{\rm d}x=\sum_{i=1}^m
{\bar x}_i p_i.
\end{equation}
Denoting ${\widetilde {\rm SER}}\left(\sqrt{{\tilde x}_i}\right)$ and ${\bar x}_i$ as $\alpha_i$ and $\beta_i$ in this case and repeating the approach for all intervals on which $f(x)>0$, we can again obtain a system of linear equations in the form of (\ref{P2}). Note that ${\hat x}_{i}\rightarrow {\hat x}_{i-1}$
as the number of  sub-intervals  $m$ increases. In this case we have ${\bar x}_i\simeq {\tilde x}_i \simeq {\hat x}_{i-1} \simeq {\hat x}_{i-1}$. Further, as ${\widetilde {\rm SER}}\left(x\right)$ is a monotonically increasing function, we can order the coefficients such that $\alpha_1\leq\alpha_2\leq\cdots\leq\alpha_n$ and
$\beta_1\leq\beta_2\leq\cdots\leq\beta_n$. Then, as $0\leq p_i\leq 1$ for all $i=1,\cdots,n$, (\ref{esys2a}) and (\ref{esys2}) indicate that $\beta_n\geq{\bar P}$.

As $f(x)$ is the optimal distribution and $b$ is the maximum of the objective function (\ref{obj}), $p_i$ with $i=1,2,\ldots,n$ in (\ref{P2}) must be a solution of the following problem
\begin{subequations}\label{P3}
\begin{align}
\max_{y_{i},i=1,\ldots,n} \!\!&\!\!&\!\!\alpha_1 y_1 +\alpha_2 y_2+\cdots +\alpha_n y_n \label{objp3} \\
\text{s.t.}\!\!&\!\!&\!\! \beta_1 y_1 +\beta_2 y_2+\cdots  +\beta_n y_n ={\bar P}\label{esys3a}\\
 \!\!&\!\!&\!\! y_1 + y_2+\cdots  + y_n =1 \label{esys3}\\
 \!\!&\!\!&\!\! 0\leq y_i\leq 1 \text{ for } i=1,2,\ldots,n\label{esys3b}
\end{align}
\end{subequations}
where $\alpha_i$ and $\beta_i$ with $i=1,2,\ldots,n$ are identical to the corresponding parameters in (\ref{P2}).
Following the KKT conditions \cite[p. 243]{Boyd}, at the maximum of the problem (\ref{P3}), we must have
\begin{equation}
 \lambda_0 =\frac{\alpha_i-\alpha_k}{\beta_i-\beta_k} \label{conds}
\end{equation}
for any $i\neq k$ with $0<p_i<1$ and $0<p_k<1$, where $\lambda_0$ is a constant (Lagrange multiplier).
Without loss of generality, let us denote $k^{\prime}$ as the first index such that $0<p_i<1$, i.e., $k^{\prime}\leq i$ for any $0<p_{k^{\prime}},p_i<1$. Similarly, we can denote $k^*$ as the last index such that $0<p_i<1$, i.e., $k^*\geq i$ for any $0<p_{k^*},p_i<1$. Clearly, we have $\beta_{k^{\prime}}\leq {\bar P}\leq \beta_{k^{*}}$, as otherwise $p_i$ with $i=1,\ldots,n$ cannot meet the constraints (\ref{esys3a})-(\ref{esys3b}).
Inserting (\ref{conds}) in (\ref{P3}) and eliminating all $\alpha_i$ and $\beta_i$ for $i\neq k^{\prime}$, the optimum
value of $b$ can be expressed as
\begin{equation}
 b= \lambda_0 {\bar P}+\alpha_{k^{\prime}}-\lambda_0\beta_{k^{\prime}}.\label{conds2}
\end{equation}
Moreover, let
\begin{equation}
 y_{k^{\prime}}\!=\!\frac{\beta_{k^*}-{\bar P}}{\beta_{k^*}-\beta_{k^\prime}};
 y_{k^{*}}\!=\!\frac{{\bar P}-\beta_{k^\prime}}{\beta_{k^*}-\beta_{k^\prime}};
 \text{ and } y_i\!=\!0 \text{ for } i\neq k^\prime, k^*
\label{valy}
\end{equation}
 and consider the following PDF ${\tilde f}(x)$
 \begin{equation}
  {\tilde f}(x)=\frac{\beta_{k^*}-{\bar P}}{\beta_{k^*}-\beta_{k^\prime}}\delta\left(x-\beta_{k^\prime}\right)+\frac{{\bar P}-\beta_{k^\prime}}{\beta_{k^*}-\beta_{k^\prime}}\delta\left(x-\beta_{k^*}\right).\label{valy2}
 \end{equation}
Inserting (\ref{valy2}) into (\ref{P1}), we obtain the same maximum value $b$ for the objective function  (\ref{obj}) as given in (\ref{conds2}) while meeting all the constraints. Therefore, there exists a PDF $f(x)$ solving the problem (\ref{P1}) with $f(x)>0$ for at most two values  of $x$. \qed

\bibliographystyle{IEEEtran}
\bibliography{Bibfile}

\end{document}